\documentclass[12pt]{article}

\usepackage[margin=1.2in]{geometry}
\usepackage{booktabs}
\usepackage{threeparttable}
\usepackage{amsmath,amssymb}
\usepackage{setspace}
\usepackage{natbib}
\usepackage{hyperref}
\usepackage{graphicx}
\usepackage{caption}
\usepackage{subcaption}
\usepackage[T1]{fontenc}
\usepackage{lmodern}
\usepackage{microtype}
\usepackage{xcolor}
\usepackage{array}
\usepackage{tabularx}
\usepackage{rotating}

\hypersetup{colorlinks=true,linkcolor=blue!60!black,citecolor=blue!60!black,
            urlcolor=blue!60!black}

\title{\textbf{Colonial Rule and Religious Change:\\
Evidence from Africa's Colonial Borders}}
\author{Hector Galindo-Silva\thanks{Department of Economics, Pontificia Universidad Javeriana. Email: \href{mailto:galindoh@javeriana.edu.co}{galindoh@javeriana.edu.co}.}}
\date{\today}

\begin{document}
\maketitle
\begin{abstract}
\noindent
The European colonization of sub-Saharan Africa drove a massive shift from indigenous religions to Christianity, yet the channels through which this transformation occurred remain poorly understood. Using a geographic regression discontinuity design at colonial borders in sub-Saharan Africa, I find that Christian adherence is substantially higher under French and Portuguese direct rule than under British indirect rule---a gap that implies a correspondingly greater persistence of traditional religions where indirect rule prevailed. Neither mission presence nor pre-colonial political centralization can account for the discontinuity. Instead, the evidence points to the disruption of the inherited social order as the key channel: where direct rule eroded rigid traditional social structures, Christianity---which bypassed hereditary boundaries---expanded to fill the void; where indirect rule preserved them, indigenous religions endured. These findings shed light on the dynamics of religious identity change and how it was shaped by colonialism.
\end{abstract}

\newpage
\section{Introduction}

Religious identities, despite their persistence, do change. Deep institutional shocks are among the most important drivers of such change \citep[for overviews, see][]{iannaccone1998introduction, iyer2016economics}. The European colonization of sub-Saharan Africa offers a particularly salient case. Around 1900, an estimated 76\% of the region's population practiced traditional religions, 14\% were Muslim, and only 9\% were Christian; by 2010, roughly 57\% identified as Christian and 29\% as Muslim, while traditional religions had declined to approximately 13\% \citep{pew2010tolerance}.

A plausible channel linking colonial rule to this religious transformation is missionary activity: there is evidence that European colonial powers significantly facilitated the expansion of Christian missions \citep{becker2022empire, jedwab2022christianity}, which in turn could have accelerated conversion. Yet the success of missions also depended on local demand for the new faith and on the institutional environment shaped by colonial administrations \citep{horton1971african, kalu2008african}. This paper studies one aspect of this broader problem: whether the type of colonial rule shaped long-run Christianization---for example, through differently supplying missions or unequally disrupting the social structures in which indigenous religions were embedded.

The sub-Saharan African experience provides a clear framework for this inquiry, as the region's colonial history is defined by a sharp divergence in governance models. British indirect rule delegated authority to recognized chiefs and preserved customary social arrangements \citep{mamdani1996citizen, bolt2025councils}. French and Portuguese direct rule, by contrast, subordinated or replaced traditional authorities as part of assimilationist projects that sought to dismantle indigenous social institutions \citep{crowder1968west, young1994the, conklin1997mission, newitt1995mozambique}.

I analyze this contrast by examining the impact of colonial rule—direct or indirect—on long-run Christian adherence relative to the retention of traditional religions. A preliminary OLS analysis using cross-sectional variation at the ethnic group level reveals a negative yet statistically non-significant correlation between indirect rule and Christian adherence. While imprecise, this negative coefficient suggests a pattern consistent with my subsequent findings: indirect rule was associated with lower Christianization---and thus greater retention of traditional religions---relative to direct rule.

Because the placement of colonial borders was likely not random in many places, I employ a geographic RDD design across divided ethnic homelands. This design focuses on eight colonial border segments that partition ethnic homelands between the two regimes.\footnote{The sample includes six British–French borders (Ghana–Togo, Nigeria–Cameroon, Ghana–Burkina Faso, Ghana–C\^{o}te d'Ivoire, Sierra Leone–Guinea, and Nigeria–Benin) and two British–Portuguese borders (Zambia–Mozambique and Zimbabwe–Mozambique). See Section~\ref{sec:strategy} for a detailed accounting.} By comparing sub-units of the same ethnic group on opposite sides of these boundaries, I am able to hold constant pre-colonial ancestral traits and isolate the specific impact of the colonial administration.

The RDD estimates, leveraging this cleaner within-homeland variation, uncover a large and significant gap consistent with the direction of the OLS findings: communities on the French and Portuguese sides of the border have Christian population shares roughly 15–17 percentage points higher than observationally similar communities on the British side. This finding, obtained via a non-parametric local linear estimator with bias-corrected robust inference \citep{calonico2014robust}, is robust to the inclusion of geographic controls, various bandwidth specifications, and homeland fixed effects. Decomposing the results by denomination, I find that the effect is concentrated in non-evangelical Christianity---suggesting that the gap reflects a deeper institutional legacy rather than mere cultural inertia.

To identify the drivers of this divergence, I evaluate several potential mechanisms. Surprisingly, the effect is not accounted for by mission presence: neither Catholic nor Protestant mission density exhibits a discontinuity at the border. The heterogeneity analysis further reveals that pre-colonial political centralization—whether measured by jurisdictional hierarchy or democratic leadership selection—does not moderate the gap either. If missions and political institutions are not the drivers, what is? I argue that the key channel was the disruption of pre-colonial social structures. Where French and Portuguese direct rule dismantled hereditary hierarchies that regulated marriage, occupation, and social mobility \citep{tamari1991caste}, Christianity—which cut across these distinctions—could fill the resulting vacuum \citep{stark2000acts}. Conversely, where British indirect rule preserved them, indigenous practices retained the social cohesion needed to endure. Consistent with this, the direct-rule advantage is significantly amplified in societies with endogamous class stratification, and the result is robust to alternative measures of hereditary hierarchy. These findings align with a growing literature showing that the effects of colonialism depended critically on how regimes interacted with pre-existing social structures \citep{mamdani1996citizen, bolt2025councils, heldring2021dissolution}.

\medskip
This paper contributes to several literatures. First, I provide the first causal evidence that colonial governance shaped religious outcomes through channels that bypass missionary infrastructure, extending the literature on the long-run effects of colonialism to the dimension of religious identity. Second, the findings advance the economics of religion \citep{iyer2016economics} by demonstrating that the expansion of world religions depends not only on the supply of religious entrepreneurs but also on the administrative environment's interaction with pre-existing social organization. In the ``religious economy'' framework, new religions succeed when they offer functional benefits that incumbents fail to provide \citep{stark2000acts, rubin2017rulers}; my results suggest that the identity of the colonial power determined whether the social structures underpinning indigenous practices survived or collapsed. Third, I complement the extensive literature on missions and African development \citep{nunn2010missionary,jedwab2022christianity,woodberry2012missionary} by showing that mission density cannot account for the colonial discontinuity in Christianization, suggesting that the institutional channel operates independently of missionary infrastructure.

More broadly, a large quantitative literature has documented lasting colonial imprints on institutions, human capital, ethnic identity, state capacity, and cultural outcomes.\footnote{See, among others, \citet{acemoglu2001colonial} on institutions, \citet{huillery2009history} and \citet{gallego2010missionaries} on human capital, \citet{Michalopoulos2016} on ethnic identity, \citet{herbst2000states} and \citet{mullercrepon2020indirect} on state capacity, and \citet{nunn2011slave} and \citet{lowes2021concessions} on cultural outcomes.} A growing strand emphasizes that colonialism's effects depended on how regimes interacted with pre-existing social structures \citep{mamdani1996citizen, bolt2025councils, heldring2021dissolution}; this paper provides quantitative evidence for that proposition in the domain of religious change.

Methodologically, my identification strategy follows \citet{Michalopoulos2016} and \citet{michalopoulos2014nations}, who introduced the within-split-homeland comparison across colonial borders.\footnote{\citet{Michalopoulos2016} apply this design to satellite light density; \citet{michalopoulos2014nations} extend it to show that national institutions affect subnational development within partitioned homelands.} Most closely related on substance are \citet{cogneau2014borders}, who document persistent colonial effects on education and religious affiliation at the Ghana--Togo border---one of my eight segments---and \citet{dupraz2019colonial}, who estimates differential effects of British and French rule on education along the Nigeria--Cameroon border.\footnote{On the mechanisms side, \citet{caicedo2019mission} uses spatial RD around Jesuit territories in South America to identify long-run mission effects; I apply analogous logic but find a null effect, suggesting that the colonial channel bypasses missionary density.} My focus on religious change as a first-order outcome of colonial domination builds on \citet{nunn2010missionary}, and the administrative interaction mechanism I identify extends the insight of \citet{bolt2025councils} to the religious landscape.

The remainder of the paper is organized as follows. Section~\ref{sec:data} describes the data. Section~\ref{sec:strategy} presents the empirical strategy. Section~\ref{sec:results} reports the main results and robustness checks. Section~\ref{sec:mechanisms} examines mechanisms. Section~\ref{sec:conclusion} concludes.

\section{Data}\label{sec:data}

The empirical analysis combines data on colonial administration, ethnic homelands, religious demography, historical missions, and geo-ecological conditions.

To identify colonial assignment, I rely on the Colonial Dates Dataset \citep[COLDAT;][]{Becker2019}, which records the identity and dates of all colonial rulers for every historically administered territory.\footnote{%
  Data available at \url{https://dataverse.harvard.edu/dataset.xhtml?persistentId=doi:10.7910/DVN/T9SDEW}.}
From this source, I construct the binary treatment indicator for British versus French/Portuguese rule. In my sample, British rule spans from 1874 in Ghana to 1899 in Nigeria; French administration ranges from the 1880s to the mandates established in 1922; and Portuguese presence in Mozambique dates to the sixteenth century, with effective administrative control consolidated in the late nineteenth century. All territories thus experienced sustained colonial governance well before the outcomes were measured.

These colonial jurisdictions are overlaid upon ethnic homeland polygons sourced from the Ancestral Characteristics of Modern Populations database  \citep{giuliano2018ancestral}.\footnote{%
  Data available at \url{https://nathannunn.arts.ubc.ca/data/}.}
These polygons are constructed from the \textit{Ethnologue: Languages of the World} \citep{lewis2009ethnologue}, which maps the global distribution of languages into mutually exclusive geographic units. These polygons are matched to the Ethnographic Atlas \citep{murdock1967ethnographic}, a global database covering over 1,100 societies with pre-colonial ethnographic characteristics such as social stratification and political organization. I use these polygons to identify ethnic homelands that were divided by colonial borders---the foundation of my identification strategy. The unit of observation is defined as the centroid of each sub-homeland polygon; since a single homeland’s territory may be split across the colonial border, this within-homeland geographic variation provides the essential identifying variation for my regression discontinuity design.

To measure religious transformation within these units, I utilize the Joshua Project (JP), a global registry of over 17,000 people groups that provides detailed data on the share of each group’s population identifying as Christian.\footnote{%
  The JP data are aggregated from a variety of sources, including national censuses, field reports from mission agencies, and global databases like Ethnologue and the World Christian Encyclopedia. These data have well-known limitations: adherence estimates are often ``ballpark’’ figures scaled to the current year from older census data; the definition of ``evangelical’’ may reflect ecclesiological rather than demographic criteria; and there is often a time lag between field changes and database updates. Despite these caveats, JP remains the most comprehensive global source of ethno-linguistic religious composition, used by \citet{nunn2010missionary} and \citet{jedwab2022christianity}, among others. Data available at \url{https://joshuaproject.net/data}.} JP reports both evangelical and non-evangelical Christian shares separately, which I use to construct two complementary metrics: the total Christian population share and the persistence of traditional religions, defined as the share of the population that has not converted to Christianity. Since all ethnic groups in my sample were primary practitioners of ethnic religions at the onset of colonization, this latter variable effectively captures the retention of indigenous beliefs.\footnote{Although Islam is a significant force in West Africa, its role in my specific sample of divided homelands appears secondary, as the primary substitution I observe is between indigenous beliefs and Christian denominations. The Joshua Project classifies all divided homelands in my sample as primary practitioners of ``Ethnic Religions,'' suggesting limited Islamization at the onset of the colonial era. To ensure my results are not driven by varying proximity to the Islamic Sahel, I use the latitude of each centroid as a proxy for Islam exposure; my findings are robust to this control. The fact that the British side shows higher traditional religious persistence rather than higher Islamization reinforces my argument that indirect rule acted as a buffer for local traditions.}

To test whether differential missionary presence accounts for the religious gap, I use georeferenced mission station data from \cite{becker2022empire}, \cite{becker2022mapping}, and \cite{becker2022polygamy}. These datasets provide a comprehensive record of Protestant and Catholic missions, offering a longitudinal perspective that addresses the selection bias inherent in static mission atlases. By computing the number of stations within a 100~km radius of each polygon centroid, I can test whether the colonial border generated a discontinuity in missionary presence or composition. These historical records are further complemented by a suite of geographic and ecological controls—including elevation, temperature, precipitation, caloric suitability, distance to the coast, absolute latitude, terrain ruggedness, and distance to the nearest navigable river—to ensure that my comparisons are not confounded by local environmental factors.\footnote{%
  The geographic controls are drawn from the following sources: elevation from the Shuttle Radar Topography Mission \citep[SRTM;][]{farr2007srtm}; temperature and precipitation from WorldClim~v2.1 \citep{fick2017worldclim}; caloric suitability from the pre-1500 Caloric Suitability Index \citep{galor2016agricultural}; distance to the coast computed as Haversine distance to the nearest coastline; terrain ruggedness from Riley's Terrain Ruggedness Index at 30 arc-second resolution \citep{nunn2012ruggedness}; and distance to the nearest navigable river from Johnston's historical map of African waterways, digitized by \citet{jedwab2022christianity}.}
 Appendix Table~\ref{tab:desc_stats} provides summary statistics for all variables.
\section{Empirical Strategy}\label{sec:strategy}

A natural starting point for the analysis is to examine the association between colonial rule and religious outcomes across the full African sample using a cross-sectional OLS framework:
\begin{equation}\label{eq:ols}
  Y_i \;=\; \alpha \;+\; \beta\,D_i \;+\; \delta' X_i
  \;+\; \phi_c \;+\; \varepsilon_i,
\end{equation}
where $Y_i$ is the Christianization outcome for ethnic homeland polygon~$i$, $D_i$ is an indicator for British colonial rule, $X_i$ is a vector of geographic and ecological controls, and $\phi_c$ denotes grid-cell fixed effects that restrict comparisons to homelands within the same broad geographic zone. Despite these controls, the OLS estimate of $\beta$ may be biased by the potentially endogenous placement of colonial borders, which likely reflected strategic competition and thus introduced systematic differences that geographic controls alone cannot fully address. These concerns motivate a more robust identification strategy.

\begin{figure}[p]
\centering
\caption{Study Areas: Colonial Borders in the Estimation Sample.}
  \includegraphics[width=0.85\textwidth]{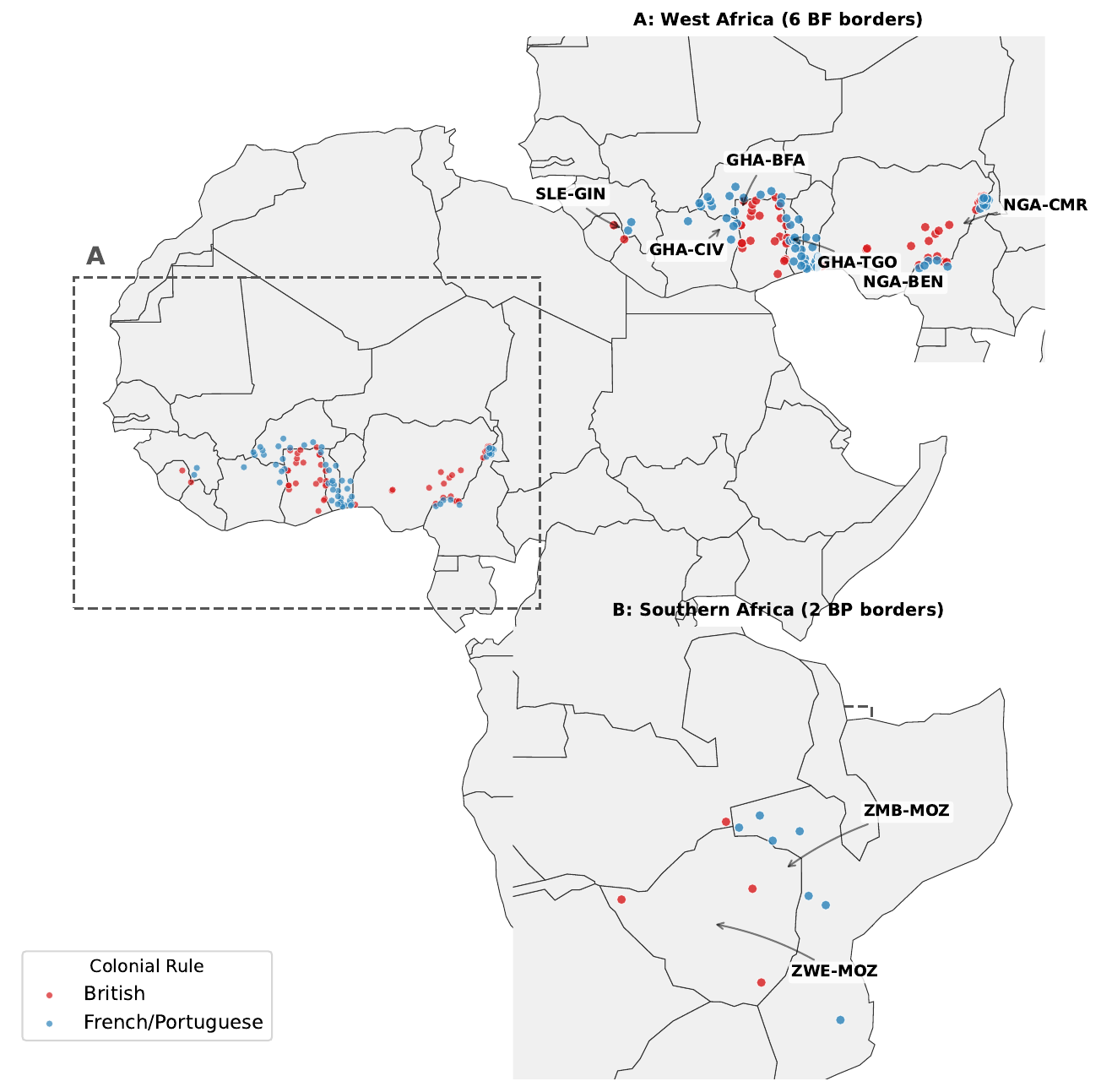}
\label{fig:africa_map}
\begin{minipage}{0.92\textwidth}
\vspace{4pt}
\scriptsize
\textit{Notes:}
The map displays the eight colonial border segments analyzed in the RDD:
six British--French (GHA--TGO, NGA--CMR, GHA--BFA, GHA--CIV, SLE--GIN, NGA--BEN)
and two British--Portuguese (ZMB--MOZ, ZWE--MOZ).
Each point represents a sub-homeland polygon centroid.
Red markers denote centroids on the British side ($r_i \geq 0$);
blue markers denote centroids on the French/Portuguese side ($r_i < 0$).
Inset~A zooms into the six West African borders; Inset~B into the two Southern African borders.
\end{minipage}
\end{figure}

To overcome these challenges, I shift my focus to a sharp geographic regression discontinuity design (RDD) that exploits the quasi-random partition of Africa. I construct a specialized estimation sample by identifying ethnic homelands that were historically divided by eight colonial border segments: six British--French (Ghana--Togo, Nigeria--Cameroon, Ghana--Burkina Faso, Ghana--C\^{o}te d'Ivoire, Sierra Leone--Guinea, and Nigeria--Benin) and two British--Portuguese (Zambia--Mozambique and Zimbabwe--Mozambique). These segments represent the exhaustive set of British versus French/Portuguese land borders in sub-Saharan Africa where the Murdock atlas records divided ethnic groups, spanning diverse ecological zones from the West African coastal forest to the Southern African plateau (see Figure~\ref{fig:africa_map}). By ensuring that the same pre-colonial group appears on both sides of these boundaries, I can compare 122 sub-homeland polygon centroids drawn from 26 divided homelands (listed in Appendix Table~\ref{tab:ethnic_groups}), using their signed geodesic distance $r_i$ to the border as the running variable.\footnote{%
  There are 20 British--French and British--Portuguese colonial land borders in Africa. Twelve are excluded: two (South Sudan--CAR, Eswatini--Mozambique) lack divided homelands entirely; six (Nigeria--Niger, Nigeria--Chad, Sudan--Chad, Sudan--CAR, Somalia--Djibouti, Kenya--Somalia) lie in Muslim-majority regions with near-zero Christian adherence on both sides; four (Malawi--Mozambique, Tanzania--Mozambique, Zambia--Angola, South Africa--Mozambique) have extensive missing data in the Joshua Project; and one (Gambia--Senegal), despite having eight divided homelands, yields only four observations with non-missing religious data, of which only one homeland (Pepel) has observations on both sides of the border---too few to contribute meaningful within-homeland variation. The remaining eight borders are all those where the RDD can be implemented.}
By construction, the historical placement of colonial borders makes selective sorting or manipulation highly unlikely. 

This design identifies the local average treatment effect (LATE) by comparing outcomes around the border cutoff ($r_i=0$), where $D_i = \mathbf{1}[r_i \geq 0]$ signifies the British side. Because I am comparing sub-units of the same ethnic homeland, this approach effectively holds constant all time-invariant pre-colonial ancestral traits, isolating the impact of the different colonial regimes. The primary results are estimated using the robust nonparametric local linear RDD estimator developed by \citet{calonico2014robust}. The treatment effect $\tau$ is recovered by solving the following weighted minimization problem:
\begin{equation}\label{eq:np}
  \min_{\alpha, \tau, \beta, \gamma} \sum_{i=1}^n \left[ Y_i - \alpha - \tau D_i - \beta r_i - \gamma (D_i \times r_i) \right]^2 K\left(\frac{r_i}{h}\right),
\end{equation}
where $K(\cdot)$ is a triangular kernel and $h$ is a data-driven MSE-optimal bandwidth. To ensure the results are not sensitive to the specific window or estimator, I complement this nonparametric approach with a parametric RDD—estimated across a range of fixed bandwidths $h \in \{75, 100, \dots, 300\}$~km—that includes homeland fixed effects ($\mu_g$) and border-specific quadratic polynomials.\footnote{The parametric specification is $Y_i = \alpha + \beta\,D_i + \sum_{b}(\gamma_{1b}\,r_i\cdot\mathbf{1}[b_i=b] + \gamma_{2b}\,r_i^2\cdot\mathbf{1}[b_i=b]) + \mu_g + \varepsilon_i$, where $b_i$ indexes the border segment. Homeland fixed effects are crucial as they absorb all time-invariant homeland-level characteristics—including language, pre-colonial institutions, and traditional religious trajectories—ensuring that my identification relies exclusively on within-homeland, cross-border variation.}

The causal interpretation of these estimates rests on the assumption that potential outcomes vary smoothly across the colonial boundary. In the context of the ``Scramble for Africa,'' this assumption is plausible: colonial borders were largely determined by European diplomats during the 1884--1885 Berlin Conference, often drawn using straight latitudinal or longitudinal marks on imprecise maps, with little regard for local ethno-linguistic distributions or terrain. From the perspective of the ethnic groups whose homelands were subsequently split, the border's location represents a plausibly exogenous shock. However, as \citet{Michalopoulos2016} note, colonial borders sometimes followed natural features such as rivers or mountain ranges, raising the concern that ecological discontinuities could confound the RDD. I address this by testing the smoothness of eight geo-ecological covariates at the boundary and using homeland fixed effects to absorb shared geographic heterogeneity. Table~\ref{tab:balance_geo} reports nonparametric balance tests: seven of the eight covariates show no significant discontinuity; the exception is temperature, which exhibits a  significant difference; nevertheless, the results are consistently robust to controlling for this and all other covariates.\footnote{The eight covariates are chosen to capture the main dimensions along which geographic confounders could operate: (i)~\textit{elevation} controls for altitude-driven differences in disease ecology, crop suitability, and accessibility; (ii)~\textit{temperature} and (iii)~\textit{precipitation} capture climatic gradients that affect both agricultural potential and missionary settlement patterns; (iv)~\textit{caloric suitability} (pre-1500) measures the agricultural endowment prior to any colonial influence; (v)~\textit{distance to the coast} proxies for exposure to trade, early colonization, and missionary access; (vi)~\textit{absolute latitude} captures north--south gradients correlated with proximity to the Islamic Sahel; (vii)~\textit{terrain ruggedness} (Riley's TRI from SRTM data) measures topographic heterogeneity, which affects colonial penetration and state capacity \citep{nunn2012ruggedness}; and (viii)~\textit{distance to the nearest navigable river} captures access to the transportation routes that facilitated both colonial administration and missionary expansion. The isolated rejection of temperature (1 of 8 covariates at the 5\% level) is consistent with the expected false positive rate.}

A related concern is whether individuals or missions migrated across the border in the post-colonial period---perhaps to seek mission schools or to avoid the more interventionist French/Portuguese administration. Several factors limit this threat. First, the unit of observation is a fixed geographic polygon (the ethnic homeland territory), so while people can move, the homeland itself remains stationary. Second, such migration would likely attenuate the estimated gap: if individuals from the British side moved to the French/Portuguese side to access missions, the Christian share on the comparison side would rise, biasing the estimate toward zero. Conversely, if individuals fled the more centralized administration, they would carry their religious identities with them, again smoothing out the discontinuity. The persistence of a large gap despite these forces suggests that the estimates capture a structural institutional legacy rather than a demographic artifact.

Beyond these core assumptions, several caveats qualify my interpretation. First, the RDD sample is small. I expanded it as far as the data allow by pooling all eight viable border segments across Africa, but the split-homeland requirement and data availability impose hard constraints. This reflects a deliberate trade-off: the within-homeland RDD sacrifices external validity for internal validity, although the main estimates are precisely estimated and the nonparametric bandwidth selection yields sufficient effective observations for inference, power for detecting smaller effects or for demanding subgroup analyses is naturally more limited. Reassuringly, the OLS estimates on the full African sample exhibit the same negative signs across all outcomes. Second, geographic spillovers across the border may attenuate the estimated gap, so my estimates are best interpreted as a conservative lower bound. Third, the colonial treatment is a ``bundled'' package of administrative and legal policies; my mechanism analysis explicitly attempts to disentangle these channels. Finally, the outcome data reflect a single contemporary cross-section, but because religious identity is highly persistent, these measures are intended to capture the cumulative, long-run legacy of colonial rule.

\section{Main Results}\label{sec:results}

Table~\ref{tab:types_christian}, Panel A presents cross-sectional OLS estimates for the full African sample, where the Christian population share is regressed on the British colonial indicator with geographic and grid-cell controls. The coefficients are $-0.9$ percentage points for the total Christian share (Column 1), $-0.4$ for the evangelical share (Column 2), and $-0.5$ for the non-evangelical share (Column 3). At first glance, these findings might suggest that colonial administration had little impact on long-run religious change. However, these cross-sectional comparisons likely suffer from substantial measurement error and omitted-variable bias that would naturally attenuate the results toward zero. Despite these identification challenges, a noteworthy feature of the results is the consistent negative sign across all outcomes—a pattern that persists and becomes statistically significant in the RDD analysis.

\begin{figure}[p]
\centering
\caption{RDD plot: \% Christian adherents at colonial borders.}
  \includegraphics[width=0.75\textwidth]{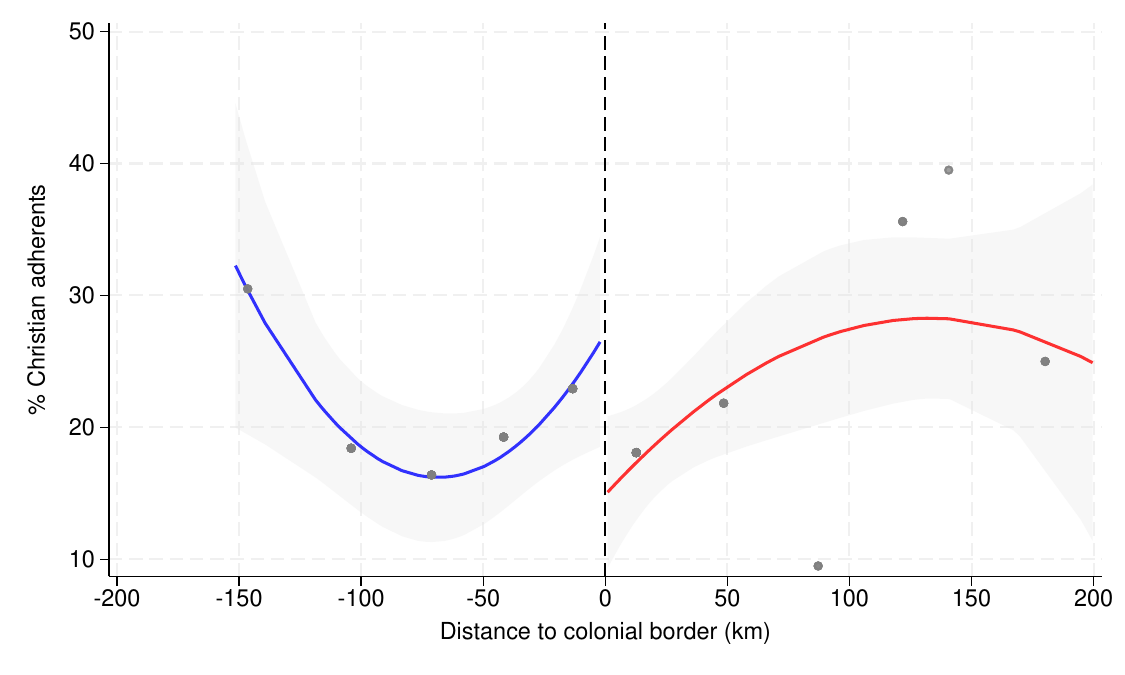}
\label{fig:rdd_scatter}
\begin{minipage}{0.92\textwidth}
\vspace{4pt}
\scriptsize
\textit{Notes:}
This figure plots the share of Christian adherents (\%, Joshua Project, circa 2010) against the signed distance to the nearest colonial border (km), where negative values indicate the French/Portuguese side and positive values indicate the British side. Each circle represents a bin-averaged outcome following the evenly spaced mimicking-variance method of \citet{calonico2014robust}. The solid lines show quadratic polynomial fits estimated separately on each side of the cutoff; shaded areas represent pointwise 95\% confidence intervals. The visible downward jump at the border indicates that communities on the British side have substantially lower Christian adherence than those on the French/Portuguese side.
\end{minipage}
\end{figure}

Once I restrict the analysis to geographically similar communities within the same ethnic homeland using my RDD framework, the negative association between British rule and Christianization becomes both larger in magnitude and statistically significant. The central findings, presented in Panel B of Table~\ref{tab:types_christian}, indicate a sharp discontinuity at the border. Specifically, the robust nonparametric estimate for the total Christian population share (Column 1) shows that communities on the British side have 16.9 percentage points fewer Christians than their counterparts on the French/Portuguese side, a result statistically significant at the 1\% level. This discontinuity is visually apparent in Figure~\ref{fig:rdd_scatter}, which shows a sharp downward jump in Christian adherence at the border as one moves from French/Portuguese to British jurisdiction.

Decomposing this aggregate result by denomination, Columns 2 and 3 of Panel B reveal that the British deficit is concentrated almost entirely in non-evangelical Christianity. The robust gap for non-evangelical adherents is -14.0 percentage points (significant at the 5\% level), accounting for approximately 83\% of the total discontinuity. In contrast, the effect on evangelical shares is much smaller and remains statistically indistinguishable from zero (-1.4 percentage points). The shift from the small, insignificant OLS coefficients in Panel A to these large, significant RDD estimates in Panel B underscores how the within-homeland comparison purges the confounds that otherwise obscure the deep institutional legacy of colonial rule.

Crucially, this deficit in Christianization on the British side provides direct, empirical evidence for the persistence of traditional religions. In my sample of ethnic groups practicing indigenous beliefs at the onset of colonization, and given the limited role of Islamization in these divided homelands (see Section~\ref{sec:data}), the variation I observe is fundamentally a substitution between traditional practices and new faiths; every percentage point of ``missing'' Christianization on the British side corresponds to a percentage point of indigenous religious retention. Consequently, the observed discontinuity of 17 percentage points suggests that British colonial rule was characterized by a higher survival rate of traditional religious systems relative to French and Portuguese rule, which appears to have more effectively facilitated the expansion of Christianity among populations with similar pre-colonial religious backgrounds.

\bigskip
These results are robust to alternative specifications and sample definitions. Appendix Table~\ref{tab:types_christian_param} presents parametric RDD estimates using a quadratic polynomial in distance and homeland fixed effects across a wide range of fixed bandwidths ($h \in \{50, 75, \dots, 300\}$~km). The coefficients remain stable in sign, ranging from $-6.3$ to $-8.5$ percentage points, and are statistically significant at most bandwidths, confirming that the discontinuity is a structural feature of the data rather than an artifact of a particular bandwidth or weighting scheme. Because the balance test (Table~\ref{tab:balance_geo}) detected a marginally significant discontinuity in temperature, I also estimate the preferred specification with all eight geographic covariates. Panel~A of Appendix Table~\ref{tab:rdd_full_vs_bf} reports these results: the robust coefficient drops from $-16.9$ to $-11.2$ percentage points but remains statistically significant at the 5\% level, confirming that the temperature imbalance does not account for the main finding. Finally, because the sample pools six British--French and two British--Portuguese borders, a natural concern is whether the Southern African segments---where ecological conditions, colonial histories, and missionary traditions differ substantially from those in West Africa---drive the main result. Panel~B of Appendix Table~\ref{tab:rdd_full_vs_bf} restricts the sample to the six British--French borders. The point estimates are similar in sign and magnitude to the full-sample baseline (Table~\ref{tab:types_christian}, Panel~B), confirming that the two British--Portuguese segments do not drive the result.

\section{Mechanisms}\label{sec:mechanisms}

Having established that direct colonial rule generated a significant advantage in Christianization---and, by extension, that British indirect rule facilitated a notably higher persistence of traditional religious systems---I now investigate the channels through which colonial identity shaped these long-run outcomes. I examine three principal channels---the spatial supply of missionary infrastructure, the interaction of colonial administrations with pre-colonial political hierarchies, and the disruption of pre-colonial social stratification---and discuss several additional possibilities.

\subsection{Missionary infrastructure}

The most immediate explanation for the direct-rule advantage in Christianization is a higher density of missionary activity on the French/Portuguese side. Christian missions are widely documented as a major vehicle for long-run religious conversion in Africa, though their success depended on local demand and the institutional environment shaped by colonial administrations \citep{nunn2010missionary, woodberry2012missionary, becker2022empire, kalu2008african}. I test this supply-side hypothesis using three indicators based on georeferenced data from \citet{becker2022empire} and \citet{jedwab2022christianity}: a binary indicator for mission presence, separate indicators for Protestant and Catholic stations, and the denominational shares of total missions within a 100~km radius of each polygon centroid.

As reported in Table~\ref{tab:rdd_missions}, the RDD estimates for all three measures are small and statistically indistinguishable from zero---the density of mission stations varies smoothly at the border. Even with similar levels of physical missionary access, the British regime facilitated a significantly higher persistence of traditional religious systems. Although this null does not definitively rule out a role for missionary supply, it substantially lowers the plausibility of this channel, at least for the sample studied here. Moreover, interacting the British treatment indicator with mission presence (any, Catholic, and Protestant separately) yields small and statistically insignificant coefficients across all bandwidths, indicating that the colonial gap is not moderated by the local presence of missions. Since the physical infrastructure of conversion appears continuous at the boundary, the direct-rule advantage plausibly stems from deeper institutional differences in how colonial administrations interfaced with local populations.

\subsection{Political centralization}

If the gap does not originate in missionary supply, an alternative possibility is that it reflects how colonial administrations interacted with pre-existing political structures. Pre-colonial political centralization has been shown to shape long-run development outcomes \citep{michalopoulos2013precolonial}, and British and French/Portuguese regimes related differently to local hierarchies: systematic evidence from colonial administrative records shows that French colonization led to the demise of roughly twice as many pre-colonial polities as British colonization \citep{mullercrepon2020continuity}. Under indirect rule, the British governed through existing chiefs, giving local political structures a mediating role over community life; under direct rule, the French and Portuguese replaced traditional leaders, bypassing these structures altogether \citep{mamdani1996citizen, bolt2025councils}. If this asymmetry matters, the colonial gap should vary with political centralization: where British administrators governed through powerful chiefs, those chiefs could have affected the adoption of Christianity---for instance, by resisting external religious influence---while on the direct-rule side, the removal of these gatekeepers could have facilitated evangelization. Among hierarchical groups, these opposing forces would amplify the gap.

I test this by interacting the British treatment indicator with two measures of pre-colonial political organization from the Ethnographic Atlas \citep{murdock1967ethnographic}: jurisdictional hierarchy beyond the local community (EA~v33), which captures administrative levels above the village, and a binary indicator of democratic leadership selection---defined, following \citet{bentzen2019power}, as election or consensus rather than hereditary succession (EA~v72). Table~\ref{tab:heterogeneity_combined}, Panels~A and~B report the results. Both interaction coefficients are small and statistically insignificant across all bandwidths, indicating that pre-colonial political centralization does not moderate the colonial gap. The mechanism appears to operate not through the formal political architecture that colonial powers co-opted or replaced, but through other dimensions of pre-colonial social organization.

\subsection{Social stratification}

Given the null result for political centralization, a natural alternative is to look beyond the formal apparatus of governance and consider the deeper social structures that organized everyday life. Many West African societies developed endogamous caste systems---hereditary, occupation-based strata with strict marriage boundaries---documented across the Manding, Wolof, Fulani, and other groups \citep{tamari1991caste}. Influential accounts of African religious change suggest that conversion may have been driven less by missionary persuasion than by the destabilization of established social worlds \citep{horton1971african, kalu2008african}.

The two colonial regimes could have destabilized these worlds asymmetrically. The French and Portuguese assimilationist projects sought to dismantle traditional social hierarchies---including customary law, hereditary chieftainship, and local status systems \citep{young1994the, conklin1997mission}. Where social stratification was more rigid, this dismantling may have created a deeper institutional vacuum, and Christianity's universalist message---which explicitly transcended hereditary social distinctions \citep{sanneh1989translating}---could have found a more receptive audience. British indirect rule, by contrast, preserved customary social arrangements, potentially maintaining the very boundaries that an egalitarian faith would otherwise dissolve. Where direct rule dismantled the social institutions that enforced those boundaries, the resulting vacuum would have amplified the demand for conversion; where indirect rule preserved them, the traditional order retained the cohesion needed to resist it.

Broader empirical evidence is consistent with both sides of this asymmetry. Indigenous religions in Africa were not free-standing belief systems but the institutional backbone of social ethics and communal cohesion \citep{kalu2008african}; where the social order was less thoroughly disrupted, communities had less reason to abandon the religious systems embedded in it \citep{horton1971african}. Where it was disrupted, however, conversion followed: across colonial Africa, early Christian converts were disproportionately drawn from marginalized social positions---ex-slaves, outcasts, and those excluded from the privileges of the traditional order---and converts from non-elite backgrounds subsequently achieved upward social mobility, displacing traditional landed chiefs \citep{jedwab2022christianity, meierselhausen2018social}. Both forces would have operated more intensely in societies with rigid, endogamous stratification, where the boundaries between groups were hardest to cross and their dismantling most consequential.

Directly testing this channel is a daunting task, as it would require measuring how each regime reshaped the social structures it encountered. I propose an indirect test based on an observable implication: if the mechanism operates through the disruption of rigid social boundaries, the colonial gap should be larger where pre-colonial stratification was more rigid. To operationalize this, I interact the treatment indicator with class stratification based on endogamy (EA~v68), a variable from the Ethnographic Atlas that captures the degree to which social strata were reproduced through endogamous marriage rules \citep{murdock1967ethnographic}. Unlike the broader class stratification measures used in the economics literature \citep[e.g., EA~v66 in][]{michalopoulos2013precolonial}, v68 isolates the endogamous dimension---the enforcement of marriage boundaries between strata---which is the defining feature of caste-like systems in West Africa and the dimension most directly threatened by Christianity's disregard for hereditary social distinctions. The variable ranges from absence of class distinctions (coded~0), through elite stratification without endogamy (coded~1), to rigid caste-like endogamy (coded~2). In the sample, 78\% of groups exhibit elite stratification without endogamy, 16\% display endogamous class boundaries, and 6\% lack class distinctions entirely.

Panel~C of Table~\ref{tab:heterogeneity_combined} reports the results. The interaction is negative and statistically significant across all bandwidths: the direct-rule advantage in Christianization is substantially larger in societies with more rigid, endogamous social stratification.\footnote{One might worry that political centralization and social stratification capture similar underlying characteristics, in which case the null for v33 and the significant result for v68 could reflect measurement quality rather than distinct channels. However, the pairwise correlation between the two variables is only 0.14 in the sample, consistent with the ethnographic literature: many politically centralized West African states (e.g., the Ashanti) had relatively open social mobility, while some decentralized societies (e.g., the Manding) maintained rigid hereditary caste systems \citep{tamari1991caste}. Similarly, this result is distinct from kinship tightness \citep{enke2019kinship}, which does not moderate the colonial gap at any bandwidth---likely because nearly all West African societies in the sample share tight kinship structures, leaving insufficient variation. Class stratification based on endogamy captures a different dimension: not horizontal cohesion within kin networks, but the vertical rigidity of boundaries \emph{between} social strata.}

This result is corroborated by an alternative measure: the presence of hereditary aristocracy (EA~v38), which captures whether a society possessed a privileged ruling class whose status was transmitted by descent. While v68 measures the rigidity of boundaries between social strata---specifically, the degree to which marriage across strata was prohibited---v38 captures the existence of a hereditary elite within the social hierarchy. Panel~D of Table~\ref{tab:heterogeneity_combined} reports these estimates. Despite being available for a smaller subsample (68 observations), the v38 interaction is negative and highly significant at the two narrowest bandwidths ($-22.9$ and $-21.9$ percentage points, both significant at the 1\% level), and remains significant at wider bandwidths. The two measures thus capture distinct but reinforcing dimensions of the social order that direct rule disrupted: hereditary elite privilege and endogamous caste boundaries.

I regard the channel discussed in this section as the most plausible explanation for the colonial religious gap reported in Section \ref{sec:results}. The evidence presented above supports it, and the result is further consistent with a broader literature showing that colonial impacts depended on how regimes interacted with pre-existing social organization \citep{acemoglu2014chiefs, lowes2021concessions, nunn2011slave}, as well as with evidence from other historical settings that the spread of Christianity was shaped by its encounter with established social hierarchies.\footnote{\citet{schulz2019the} and \citet{henrich2020weirdest} document how the medieval Western Church systematically dissolved intensive kin-based structures---extending marriage prohibitions out to sixth cousins---to break the clan loyalties that sustained pre-Christian religions; the regions that resisted longest, such as southern Italy, retained the strongest kinship-based social organization into the modern era. In colonial India, Dalit outcaste communities converted in mass movements---most famously the Ongole revival of 1877--78---explicitly seeking to escape the ritual pollution and occupational restrictions of the caste system \citep[see][]{iyer2016economics}. In Korea, rapid Christianization in the early twentieth century followed the collapse of the Confucian social order under Japanese colonial rule.} However, I recognize that this mechanism may not be the only one. I examine several alternative mechanisms in the next section.

\subsection{Discussion of alternative channels}

Several other channels may have operated alongside social stratification and deserve consideration, even if they are harder to test directly. One possibility is that pre-colonial theological proximity to Christianity---for instance, the presence of a High God concept (EA~v34)---facilitated conversion where colonial rule disrupted traditional institutions. A difficulty in testing this hypothesis is that the variable suffers from extensive missing data, which prevents a reliable assessment. With the available observations, the interaction estimates are statistically insignificant across all bandwidths. Additionally, an independent theological channel seems improbable: rigid stratification systems often embedded religious authority within the caste hierarchy, suggesting that any effect of theological proximity could have operated through the same social structures documented above.

A second concern is that the gap is fundamentally a denominational artifact: since the effect is concentrated in non-evangelical (predominantly Catholic) Christianity, one might argue that it simply reflects the greater presence of Catholic missions in French/Portuguese colonies and Protestant missions in British ones. This interpretation cannot be definitively ruled out with the available data, as no French--Portuguese borders with divided homelands exist that would allow a direct comparison between two Catholic colonial powers. However, several pieces of evidence reduce its plausibility. One natural channel through which denomination could matter is differential mission supply---more Catholic stations on the direct-rule side, more Protestant ones on the British side. Yet the RDD estimates show no discontinuity in Catholic mission density at the border, which is inconsistent with a purely supply-side denominational explanation.\footnote{A more speculative possibility is that Catholic and Protestant missions differed in their \emph{effectiveness} at converting populations, even conditional on similar physical presence. Catholic missions, with their emphasis on institutional integration and communal ritual, may have been particularly effective at substituting for the collective functions previously served by traditional social structures \citep{hastings1994church}. However, this remains conjectural; disentangling effectiveness from institutional channels would require data not currently available.} Additionally, the baseline sample includes two British--Portuguese border segments, where the direct-rule side was governed by a predominantly Catholic power other than France; the British deficit persists along these segments, suggesting that the result is not an artifact of specifically French denominational composition. Nonetheless, a fully convincing separation of the institutional and denominational channels would require variation that the current data cannot provide.

Other alternatives relate to geographic and institutional factors. Differential investment in transportation infrastructure could have facilitated missionary movement, though any such discontinuity would likely be captured by the balance tests on economic accessibility. Proximity to the Islamic Sahel could have influenced receptivity to Christianity; however, the colonial effect remains robust to controlling for latitude as a proxy for Islam exposure, suggesting that the British disadvantage is not merely a reflection of varying distances to Islamic centers. A related channel is education: in the inter-war period, schools became the primary vehicle for mass Christianization across sub-Saharan Africa, with communities actively demanding mission schools as instruments of social mobility \citep{kalu2008african}. Since French/Portuguese and British colonies adopted different educational policies---French centralized curricula aimed at cultural assimilation, while British administrations delegated schooling largely to missions---differential educational exposure could contribute to the religious gap. However, in the absence of a discontinuity in mission presence itself, any such educational channel may operate through the same institutional vacuum created by direct rule.

\section{Conclusion}\label{sec:conclusion}

This paper documents a robust 17-percentage-point discontinuity in Christianization at sub-Saharan colonial borders, with significantly higher shares on the French/Portuguese side. Mechanism analysis suggests that differential mission presence or pre-colonial political centralization do not appear to drive the gap; instead, the direct-rule advantage is significantly amplified in societies with rigid, endogamous class stratification. This suggests that whereas French and Portuguese direct rule dismantled hereditary boundaries—creating an institutional vacuum that Christianity’s universalist message filled—British indirect rule preserved the social infrastructure that favored traditional religious retention. This interpretation aligns with literature showing that colonial impacts were mediated by pre-existing social organization \citep{acemoglu2014chiefs, lowes2021concessions, nunn2011slave} and that Christianity’s spread depended on its capacity to reshape established social hierarchies \citep{schulz2019the, henrich2020weirdest}.

These results have broader implications for the economics of religion and the study of cultural persistence. They suggest that the expansion of world religions may be shaped not only by the supply of missionary infrastructure but also by the administrative environment's interaction with pre-existing social organization. The finding that social stratification---rather than political hierarchy---moderates the colonial effect highlights that the relevant margin of institutional disruption may have operated at the level of everyday social life: marriage norms, occupational boundaries, and inherited status. The persistence of tradition on the British side is not merely a sign of inertia but may reflect the preservation of the social fabric within which indigenous religious practices were embedded.

Limitations of this study provide a roadmap for future research. Specifically, mechanisms are inferred rather than directly observed; individual-level survey data could help bridge this gap. Moreover, the concentration of results in non-evangelical denominations makes isolating institutional from denominational effects challenging, as data constraints preclude a direct comparison between French and Portuguese colonial styles. Finally, while these findings offer a robust account of the colonial legacy in sub-Saharan Africa, whether similar mechanisms operated in other post-colonial settings---such as South and Southeast Asia, where colonial powers also encountered rigid social hierarchies---remains an open question for future comparative research.

\clearpage

\begin{table}[p]
\centering
\caption{Balance Test: Geographic and Ecological Covariates}
\label{tab:balance_geo}
\scriptsize
\setlength{\tabcolsep}{6pt}
\begin{tabular}{@{} l cccccccc @{}}
\toprule
 & (1) & (2) & (3) & (4) & (5) & (6) & (7) & (8) \\
 & Elev. & Temp. & Precip. & Caloric & Dist. & Abs. & Terrain & Dist. \\
 & (m) & ($^{\circ}$C) & (mm/yr) & Suit. & Coast & Lat. & Rugged. & River \\
\midrule
Conventional & 34.097 &  2.105$^{**}$ & 43.735 & $-$84.967 & 60.801 & $-$ 0.459 & 15.965 & 29.071 \\
  & (97.832) & ( 0.974) & (155.925) & (73.291) & (110.426) & ( 1.238) & (77.030) & (42.362) \\
Robust &  5.795 &  2.582$^{**}$ & 21.461 & $-$108.895 & 78.981 & $-$ 0.535 & $-$ 1.299 & 42.828 \\
  & (111.518) & ( 1.099) & (183.693) & (84.748) & (127.473) & ( 1.497) & (94.468) & (48.397) \\
\addlinespace[2pt]
BW (km) &  70.8 &  40.5 & 103.9 &  68.5 &  96.6 &  76.6 &  59.9 &  66.5 \\
Eff.\ observations & 67 & 51 & 84 & 67 & 79 & 71 & 62 & 67 \\
\bottomrule
\end{tabular}
\begin{minipage}{0.95\textwidth}
\vspace{6pt}
\scriptsize
\textit{Notes:}
This table tests for discontinuities in eight pre-determined geographic and ecological covariates at the colonial border. Each column uses a different covariate as the dependent variable in a nonparametric sharp RDD with local linear regression, a triangular kernel, and MSE-optimal bandwidth (Calonico, Cattaneo \& Titiunik 2014). \textit{Conventional} rows show the local linear point estimate with conventional standard errors; \textit{Robust} rows show the bias-corrected estimate with robust standard errors. The running variable is the signed distance to the nearest colonial border (km), with positive values on the British side. The absence of significant discontinuities (with the exception of temperature) supports the assumption that the colonial boundary does not coincide with a geographic break. 
$^{***}p<0.01$, $^{**}p<0.05$, $^{*}p<0.1$, $^{\dagger}p<0.15$.
\end{minipage}
\end{table}
}

\begin{table}[p]
\centering
\caption{Colonial Rule and Types of Christianization}
\label{tab:types_christian}
\footnotesize
\begin{tabular}{@{} l ccc @{}}
\toprule
 & (1) & (2) & (3) \\
 & Total & Evangelical & Non-Evangelical \\
 & Adherents &  & (Catholic/Other) \\
\midrule
\multicolumn{4}{c}{\textit{Panel A: OLS}} \\
\addlinespace[2pt]
$D_i$ (British) & $-$ 0.874 & $-$ 0.394 & $-$ 0.480 \\
  & ( 1.772) & ( 0.742) & ( 1.633) \\
\addlinespace[2pt]
Observations & 578 & 578 & 578 \\
$R^2$ & 0.440 & 0.396 & 0.442 \\
\midrule
\multicolumn{4}{c}{\textit{Panel B: RDD (nonparametric)}} \\
\addlinespace[2pt]
Conventional & $-$14.834$^{***}$ & $-$ 1.007 & $-$12.107$^{**}$ \\
  & ( 4.812) & ( 1.942) & ( 4.846) \\
Robust & $-$16.857$^{***}$ & $-$ 1.413 & $-$13.976$^{**}$ \\
  & ( 5.673) & ( 2.375) & ( 5.689) \\
\addlinespace[2pt]
BW (km) &  48.7 &  74.1 &  54.7 \\
Eff.\ observations & 57 & 70 & 59 \\
\bottomrule
\end{tabular}
\begin{minipage}{0.92\textwidth}
\vspace{6pt}
\scriptsize
\textit{Notes:}
This table estimates the effect of British colonial rule on Christian adherence using three outcome variables from the Joshua Project (circa 2010): total Christian adherents (column~1), Evangelical adherents (column~2), and non-Evangelical Christians, predominantly Catholic (column~3). Panel~A reports cross-sectional OLS estimates across 578 ethnic homeland polygons in British, French, and Portuguese colonies in Africa, with geographic grid-cell fixed effects (2$^{\circ}$ lat $\times$ 3$^{\circ}$ lon) and geographic controls; standard errors are clustered at the country level. Panel~B reports nonparametric sharp RDD estimates based on local linear regression with a triangular kernel and MSE-optimal bandwidth (Calonico, Cattaneo \& Titiunik 2014). \textit{Conventional} rows show the local linear point estimate with conventional standard errors; \textit{Robust} rows show the bias-corrected estimate with robust standard errors. The running variable is the signed distance to the nearest colonial border (km), with positive values on the British side. The RDD sample comprises 26 divided ethnic homelands across eight border segments (6 British--French, 2 British--Portuguese; 122 observations). 
$^{***}p<0.01$, $^{**}p<0.05$, $^{*}p<0.1$, $^{\dagger}p<0.15$.
\end{minipage}
\end{table}
}

\begin{table}[p]
\centering
\caption{Mission Composition at the Colonial Border}
\label{tab:rdd_missions}
\footnotesize
\setlength{\tabcolsep}{4pt}
\begin{tabular}{@{} l ccccc @{}}
\toprule
 & (1) & (2) & (3) & (4) & (5) \\
 & Any & Catholic & Protestant & Cath.\ Share & Prot.\ Share \\
\midrule
Conventional &  0.231 &  0.256 &  0.126 &  0.201 &  0.040 \\
  & ( 0.178) & ( 0.179) & ( 0.105) & ( 0.164) & ( 0.066) \\
Robust &  0.258 &  0.294 &  0.147 &  0.236 &  0.034 \\
  & ( 0.213) & ( 0.211) & ( 0.127) & ( 0.191) & ( 0.080) \\
\addlinespace[2pt]
BW (km) &  54.5 &  50.3 &  44.9 &  54.1 &  42.8 \\
Eff.\ observations & 59 & 59 & 55 & 59 & 53 \\
\bottomrule
\end{tabular}
\begin{minipage}{0.95\textwidth}
\vspace{6pt}
\scriptsize
\textit{Notes:}
This table estimates the effect of British colonial rule on the composition of Christian missions near ethnic homelands. Columns~(1)--(3) use binary indicators for the presence of any mission, a Catholic mission, or a Protestant mission within 100~km of the homeland centroid. Columns~(4)--(5) measure the Catholic and Protestant shares of total missions (set to zero when no missions are present). All estimates are nonparametric sharp RDD based on local linear regression with a triangular kernel and MSE-optimal bandwidth (Calonico, Cattaneo \& Titiunik 2014). \textit{Conventional} rows show the local linear point estimate with conventional standard errors; \textit{Robust} rows show the bias-corrected estimate with robust standard errors. All estimates include eight geographic covariates (elevation, temperature, precipitation, caloric suitability, distance to coast, absolute latitude, terrain ruggedness, and distance to navigable river). The running variable is the signed distance to the nearest colonial border (km). A  entry indicates insufficient variation in the outcome within the optimal bandwidth. Sample: 8 borders (6 British--French, 2 British--Portuguese), 26 divided homelands, 122 observations. 
$^{***}p<0.01$, $^{**}p<0.05$, $^{*}p<0.1$, $^{\dagger}p<0.15$.
\end{minipage}
\end{table}
}

\begin{table}[p]
\centering
\renewcommand{\arraystretch}{0.94}
\caption{Heterogeneous Effects: Pre-colonial Institutions and Social Structures}
\label{tab:heterogeneity_combined}
\footnotesize
\setlength{\tabcolsep}{8pt}
\begin{tabular}{@{} l c cc c @{}}
\toprule
 &  & \multicolumn{2}{c}{Interaction specification} & \\
\cmidrule(lr){3-4}
 &  (1)   & (2) & (3) & \\
 & Baseline & $D_i$ (British) & $D_i \times X_g$ & $N$ \\
\midrule
\multicolumn{5}{c}{\textit{Panel A: Jurisdictional Hierarchy (EA v33)}}
 \\
\addlinespace[1pt]
$h=75$ km & $-$  6.793 & $-$  5.788 & $-$  7.712 & 70 \\
  & ( 5.002) & ( 5.255) & ( 7.849) & \\
$h=100$ km & $-$  6.622 & $-$  6.643 & $-$  0.732 & 83 \\
  & ( 4.635) & ( 4.631) & ( 7.977) & \\
$h=150$ km & $-$  9.064$^{**}$ & $-$  7.666 &   7.170 & 99 \\
  & ( 4.560) & ( 5.165) & ( 6.393) & \\
$h=200$ km & $-$  9.349$^{**}$ & $-$  7.745 &   6.969 & 105 \\
  & ( 4.431) & ( 4.756) & ( 6.289) & \\
\midrule
\multicolumn{5}{c}{\textit{Panel B: Democratic Leadership Selection (EA v72)}}
 \\
\addlinespace[1pt]
$h=75$ km & $-$  8.178 & $-$  2.296 &  12.616 & 56 \\
  & ( 6.600) & (12.593) & (20.206) & \\
$h=100$ km & $-$ 10.035$^{**}$ & $-$  9.622 &   1.626 & 68 \\
  & ( 4.797) & ( 6.035) & (11.722) & \\
$h=150$ km & $-$ 10.979$^{***}$ & $-$ 10.315$^{*}$ &   2.121 & 81 \\
  & ( 4.074) & ( 5.659) & (11.554) & \\
$h=200$ km & $-$ 11.068$^{***}$ & $-$ 11.130$^{**}$ & $-$  0.205 & 85 \\
  & ( 3.834) & ( 4.941) & (10.103) & \\
\midrule
\multicolumn{5}{c}{\textit{Panel C: Class Stratification: Endogamy (EA v68)}}
 \\
\addlinespace[1pt]
$h=75$ km & $-$  6.793 & $-$  7.935 & $-$ 12.311$^{***}$ & 70 \\
  & ( 5.002) & ( 5.150) & ( 4.158) & \\
$h=100$ km & $-$  6.622 & $-$  8.067$^{*}$ & $-$  9.681$^{***}$ & 83 \\
  & ( 4.635) & ( 4.393) & ( 3.724) & \\
$h=150$ km & $-$  9.064$^{**}$ & $-$ 10.131$^{**}$ & $-$ 14.225$^{***}$ & 99 \\
  & ( 4.560) & ( 4.596) & ( 4.271) & \\
$h=200$ km & $-$  9.349$^{**}$ & $-$ 10.360$^{**}$ & $-$ 12.067$^{**}$ & 105 \\
  & ( 4.431) & ( 4.422) & ( 4.729) & \\
\midrule
\multicolumn{5}{c}{\textit{Panel D: Hereditary Aristocracy (EA v38)}}
 \\
\addlinespace[1pt]
$h=75$ km & $-$  9.106 & $-$ 20.771$^{***}$ & $-$ 22.893$^{***}$ & 47 \\
  & (10.386) & ( 6.952) & ( 6.783) & \\
$h=100$ km & $-$ 10.853 & $-$ 23.959$^{***}$ & $-$ 21.918$^{***}$ & 51 \\
  & ( 7.077) & ( 7.744) & ( 7.810) & \\
$h=150$ km & $-$ 16.014$^{**}$ & $-$ 24.542$^{***}$ & $-$ 12.618$^{*}$ & 59 \\
  & ( 6.707) & ( 6.143) & ( 7.661) & \\
$h=200$ km & $-$ 15.899$^{**}$ & $-$ 24.597$^{***}$ & $-$ 12.598$^{*}$ & 60 \\
  & ( 6.269) & ( 6.083) & ( 7.609) & \\
\bottomrule
\end{tabular}
\vspace{0.3em}
\begin{minipage}{\linewidth}
\scriptsize
\textit{Notes:}
 This table examines whether the effect of British colonial rule on Christian adherence varies with pre-colonial institutional characteristics. Each panel interacts the British treatment indicator ($D_i$) with a different ethnographic variable ($X_g$, demeaned at its sample mean). Column~(1) reports the baseline RDD estimate without the interaction; column~(2) shows the main effect of $D_i$ in the interacted specification; and column~(3) shows the interaction $D_i \times X_g$. All specifications include homeland and border fixed effects, border-specific quadratic polynomials in distance, eight geographic covariates (elevation, temperature, precipitation, caloric suitability, distance to coast, absolute latitude, terrain ruggedness, and distance to navigable river), and standard errors clustered at the homeland level (in parentheses). The ethnographic variables, from Murdock's (1967) \textit{Ethnographic Atlas}, are: EA~v33, jurisdictional hierarchy beyond the local community (1\,=\,no levels, 2\,=\,petty chiefdoms, 3\,=\,larger chiefdoms); EA~v72, democratic leadership selection, coded following Bentzen, Hariri \& Robinson (2019) (1\,=\,election or consensus, 0\,=\,hereditary or seniority; observations with no headman office or missing data excluded); EA~v68, class stratification based on endogamy (0\,=\,absent, 1\,=\,elite without endogamy, 2\,=\,endogamous/caste-like); and EA~v38, hereditary aristocracy (1\,=\,absent, 2\,=\,present; available for a subset of 68 observations). Sample: 8 borders (6 British--French, 2 British--Portuguese), 26 divided homelands, 122 observations (except Panel~D).
 $^{*}p<0.10$; $^{**}p<0.05$; $^{***}p<0.01$.
\end{minipage}
\end{table}
}

\clearpage
\bibliographystyle{apalike}
\bibliography{colonialreligion}

\clearpage
\appendix
\setcounter{table}{0}
\renewcommand{\thetable}{A.\arabic{table}}
\section{Appendix Tables}\label{app:tables}

\begin{table}[htbp]
\centering
\caption{Descriptive Statistics: Extended RDD Estimation Sample}
\label{tab:desc_stats}
\footnotesize
\begin{tabular}{l ccccc}
\toprule
Variable & Mean & SD & Min & Max & N \\
\midrule
\multicolumn{6}{c}{\textit{Panel A: Outcomes}} \\
\addlinespace[1pt]
~~\% Christian adherents & 20.4 & 13.0 & 0.3 & 45.0 & 122 \\
~~\% Evangelical & 4.9 & 3.9 & 0.0 & 20.0 & 122 \\
~~\% Non-Evangelical Christian & 15.5 & 11.6 & 0.0 & 44.7 & 122 \\
\midrule
\multicolumn{6}{c}{\textit{Panel B: Mission variables}} \\
\addlinespace[1pt]
~~Any mission (100 km) & 0.607 & 0.491 & 0.000 & 1.000 & 122 \\
~~Any Catholic mission (100 km) & 0.484 & 0.502 & 0.000 & 1.000 & 122 \\
~~Any Protestant mission (100 km) & 0.385 & 0.489 & 0.000 & 1.000 & 122 \\
~~Catholic share of missions & 0.379 & 0.440 & 0.000 & 1.000 & 122 \\
~~Protestant share of missions & 0.228 & 0.366 & 0.000 & 1.000 & 122 \\
\midrule
\multicolumn{6}{c}{\textit{Panel C: Treatment and running variable}} \\
\addlinespace[1pt]
~~British side ($D_i$) & 0.459 & 0.500 & 0.000 & 1.000 & 122 \\
~~Distance to border (km) & -6.3 & 136.0 & -381.8 & 625.6 & 122 \\
\midrule
\multicolumn{6}{c}{\textit{Panel D: Geographic and ecological covariates}} \\
\addlinespace[1pt]
~~Elevation (m) & 384 & 296 & 1 & 1450 & 122 \\
~~Temperature ($^{\circ}$C) & 26.1 & 1.9 & 18.5 & 28.4 & 122 \\
~~Precipitation (mm/yr) & 1144 & 364 & 599 & 2640 & 122 \\
~~Caloric suitability & 1197 & 187 & 826 & 2145 & 122 \\
~~Distance to coast (km) & 442 & 263 & 18 & 1035 & 122 \\
~~Absolute latitude (degrees) & 9.72 & 3.09 & 5.94 & 22.78 & 122 \\
~~Terrain ruggedness (TRI) & 104.3 & 149.1 & 2.4 & 929.6 & 122 \\
~~Distance to navigable river (km) & 173 & 93 & 0 & 416 & 122 \\
\midrule
\multicolumn{6}{c}{\textit{Panel E: Ethnographic (Murdock 1967)}} \\
\addlinespace[1pt]
~~Jurisd.\ hierarchy beyond local (EA v33) & 2.01 & 0.92 & 1.00 & 4.00 & 122 \\
~~Class stratification: endogamy (EA v68) & 1.02 & 0.47 & 0.00 & 2.00 & 122 \\
~~Democratic leadership (EA v72, BHR) & 0.50 & 0.50 & 0.00 & 1.00 & 96 \\
~~Hereditary aristocracy (EA v38) & 1.81 & 1.25 & 1.00 & 5.00 & 68 \\
\bottomrule
\end{tabular}
\vspace{0.3em}
\begin{minipage}{\linewidth}
\scriptsize
\textit{Notes:} 
This table summarizes the variables used in the analysis for the RDD estimation sample (122 observations drawn from 26 divided ethnic homelands across 8 colonial borders: 6 British--French and 2 British--Portuguese). $D_i = 1$ if the polygon lies on the British side; the comparison group includes French and Portuguese colonies. Outcome variables measure the share of Christian adherents from the Joshua Project (circa 2010). Mission data come from the \textit{World Missionary Atlas} (WMA, 1925, Protestant) and the \textit{Atlas Hierarchicus} (AH, 1929, Catholic), both digitized by Becker (2022). 
Geographic covariates are drawn from SRTM (elevation), WorldClim v2.1 (temperature, precipitation), Galor and \"{O}zak (2016, caloric suitability), and additional standard sources (distance to coast, absolute latitude, terrain ruggedness, distance to navigable river). 
Ethnographic variables come from Murdock's (1967) \textit{Ethnographic Atlas}: v33 measures jurisdictional hierarchy beyond the local community (1 = none, 2 = petty chiefdoms, 3 = larger chiefdoms); v68 captures class stratification based on endogamy (0 = absent, 1 = elite without endogamy, 2 = endogamous/caste); v72 indicates democratic leadership selection, coded following Bentzen, Hariri \& Robinson (2019) (1 = election or consensus, 0 = hereditary or seniority; observations with no headman office or missing data are excluded); and v38 captures the presence of hereditary aristocracy (1 = absent, 2 = present; available for 68 observations).
\end{minipage}
\end{table}
}

\begin{table}[htbp]
\centering\footnotesize
\caption{Ethnic Homelands in the Split-Homeland Sample \label{tab:ethnic_groups}}
\begin{tabular}{llrl}
\toprule
 & (1) & (2) & (3) \\
Ethnic Group (name) & Border & Min. dist. (km) & Enters at \\
\midrule
Adele & GHA--TGO &   9.9 & $\geq 50$~km \\
Ginyanga & GHA--TGO &  16.6 & $\geq 50$~km \\
Konkomba & GHA--TGO &  19.3 & $\geq 50$~km \\
Nawdm & GHA--TGO &  27.4 & $\geq 50$~km \\
Ntcham & GHA--TGO &   0.9 & $\geq 50$~km \\
Wudu & GHA--TGO &  14.1 & $\geq 50$~km \\
\addlinespace
Bana & NGA--CMR &  36.1 & $\geq 50$~km \\
Glavda & NGA--CMR &   2.1 & $\geq 50$~km \\
Iceve-Maci & NGA--CMR & 108.7 & $\geq 150$~km \\
Jiru & NGA--CMR &  57.7 & $\geq 75$~km \\
Mofu-Gudur & NGA--CMR &   1.6 & $\geq 50$~km \\
Nigeria Mambila & NGA--CMR &   3.5 & $\geq 50$~km \\
Tigon Mbembe & NGA--CMR &  12.5 & $\geq 50$~km \\
\addlinespace
Bissa & GHA--BFA &   3.2 & $\geq 50$~km \\
Kasem & GHA--BFA &  18.0 & $\geq 50$~km \\
Malba Birifor & GHA--BFA &   7.5 & $\geq 50$~km \\
Moba & GHA--BFA &   7.1 & $\geq 50$~km \\
Winye & GHA--BFA &  10.7 & $\geq 50$~km \\
\addlinespace
Bondoukou Kulango & GHA--CIV &  33.1 & $\geq 50$~km \\
Nafaanra & GHA--CIV &  11.1 & $\geq 50$~km \\
\addlinespace
Kuranko & SLE--GIN &  43.3 & $\geq 50$~km \\
Northern Kissi & SLE--GIN &   7.9 & $\geq 50$~km \\
\addlinespace
Akpes & NGA--BEN &  66.1 & $\geq 75$~km \\
\addlinespace
Kunda & ZMB--MOZ &  12.1 & $\geq 50$~km \\
\addlinespace
Tewe & ZWE--MOZ &   4.8 & $\geq 50$~km \\
Tswa & ZWE--MOZ & 117.9 & $\geq 150$~km \\
\bottomrule
\end{tabular}
\vspace{4pt}
\begin{minipage}{0.92\textwidth}
\scriptsize
\textit{Notes:} This table lists the ethnic homelands (Murdock, 1959) 
included in the RDD estimation sample. The split-homeland restriction 
requires at least one sub-homeland polygon centroid on each side of a 
colonial border with non-missing centroid-to-border distance. 
Total: 26 homelands across eight border segments 
(6 British--French, 2 British--Portuguese; 122 observations).
 \textit{Min.\ dist.}~= minimum absolute centroid-to-border distance (km)
 across all polygon centroids of that homeland (determines at which
 bandwidth the homeland first contributes observations to the RDD).
 Borders: GHA--TGO = Ghana--Togo; NGA--CMR = Nigeria--Cameroon;
 GHA--BFA = Ghana--Burkina Faso; GHA--CIV = Ghana--C\^{o}te d'Ivoire;
 SLE--GIN = Sierra Leone--Guinea; NGA--BEN = Nigeria--Benin;
 ZMB--MOZ = Zambia--Mozambique; ZWE--MOZ = Zimbabwe--Mozambique.
\end{minipage}
\end{table}
}

\begin{table}[p]
\centering
\caption{Parametric RDD: Types of Christianization by Bandwidth}
\label{tab:types_christian_param}
\scriptsize
\begin{tabular}{@{} l ccc @{}}
\toprule
 & (1) & (2) & (3) \\
 & Total & Evangelical & Non-Evangelical \\
 & Adherents & & (Catholic/Other) \\
\midrule
\multicolumn{4}{c}{\textit{Panel A: RDD ($h = 75$ km)}} \\
\addlinespace[1pt]
$D_i$ (British) & $-$ 7.320$^{**}$ & $-$ 1.776 & $-$ 5.544$^{\dagger}$ \\
  & ( 3.552) & ( 2.228) & ( 3.447) \\
Observations & 70 & 70 & 70 \\
$R^2$ & 0.622 & 0.537 & 0.600 \\
\midrule
\multicolumn{4}{c}{\textit{Panel B: RDD ($h = 100$ km)}} \\
\addlinespace[1pt]
$D_i$ (British) & $-$ 8.074$^{*}$ & $-$ 2.099 & $-$ 5.976$^{\dagger}$ \\
  & ( 4.339) & ( 1.966) & ( 4.119) \\
Observations & 83 & 83 & 83 \\
$R^2$ & 0.577 & 0.493 & 0.567 \\
\midrule
\multicolumn{4}{c}{\textit{Panel C: RDD ($h = 150$ km)}} \\
\addlinespace[1pt]
$D_i$ (British) & $-$ 8.511$^{**}$ & $-$ 1.377 & $-$ 7.134$^{*}$ \\
  & ( 3.910) & ( 1.696) & ( 3.750) \\
Observations & 99 & 99 & 99 \\
$R^2$ & 0.627 & 0.543 & 0.607 \\
\midrule
\multicolumn{4}{c}{\textit{Panel D: RDD ($h = 200$ km)}} \\
\addlinespace[1pt]
$D_i$ (British) & $-$ 8.043$^{**}$ & $-$ 1.623 & $-$ 6.420$^{*}$ \\
  & ( 3.762) & ( 1.510) & ( 3.453) \\
Observations & 105 & 105 & 105 \\
$R^2$ & 0.609 & 0.527 & 0.585 \\
\midrule
\multicolumn{4}{c}{\textit{Panel E: RDD ($h = 250$ km)}} \\
\addlinespace[1pt]
$D_i$ (British) & $-$ 6.449$^{\dagger}$ & $-$ 1.501 & $-$ 4.948 \\
  & ( 4.231) & ( 1.473) & ( 3.642) \\
Observations & 113 & 113 & 113 \\
$R^2$ & 0.559 & 0.507 & 0.541 \\
\midrule
\multicolumn{4}{c}{\textit{Panel F: RDD ($h = 300$ km)}} \\
\addlinespace[1pt]
$D_i$ (British) & $-$ 6.425$^{\dagger}$ & $-$ 1.649 & $-$ 4.776 \\
  & ( 4.170) & ( 1.495) & ( 3.570) \\
Observations & 117 & 117 & 117 \\
$R^2$ & 0.573 & 0.520 & 0.546 \\
\bottomrule
\end{tabular}
\begin{minipage}{0.92\textwidth}
\vspace{6pt}
\scriptsize
\textit{Notes:}
This table reports parametric RDD estimates of the effect of British colonial rule on Christian adherence across multiple bandwidths. The three outcome variables, from the Joshua Project (circa 2010), are: total Christian adherents (column~1), Evangelical adherents (column~2), and non-Evangelical Christians---predominantly Catholic, Orthodox, and independent churches (column~3). $D_i = 1$ if the polygon lies on the British side of a colonial border; the comparison group includes French and Portuguese colonies. Each panel restricts the sample to observations within the stated bandwidth. All specifications include homeland and border fixed effects, a quadratic border-specific polynomial in distance, and standard errors clustered at the homeland level (in parentheses). Sample: 8 borders (6 British--French, 2 British--Portuguese), 26 divided homelands, 122 observations. 
$^{***}p<0.01$, $^{**}p<0.05$, $^{*}p<0.1$, $^{\dagger}p<0.15$.
\end{minipage}
\end{table}
}

\begin{table}[p]
\centering
\caption{Robustness Checks}
\label{tab:rdd_full_vs_bf}
\footnotesize
\begin{tabular}{@{} l ccc @{}}
\toprule
 & (1) & (2) & (3) \\
 & Total & Evangelical & Non-Evangelical \\
 & Adherents &  & (Catholic/Other) \\
\midrule
\multicolumn{4}{c}{\textit{Panel A: With geographic covariates (8 borders)}} \\
\addlinespace[2pt]
Conventional & $-$10.092$^{**}$ & $-$ 2.156 & $-$ 6.575$^{*}$ \\
  & ( 3.999) & ( 2.036) & ( 3.387) \\
Robust & $-$11.156$^{**}$ & $-$ 2.663 & $-$ 7.099$^{*}$ \\
  & ( 5.092) & ( 2.421) & ( 4.245) \\
\addlinespace[2pt]
BW (km) &  47.1 &  45.2 &  52.3 \\
Eff.\ observations & 57 & 56 & 59 \\
\midrule
\multicolumn{4}{c}{\textit{Panel B: British--French only (6 borders)}} \\
\addlinespace[2pt]
Conventional & $-$ 7.993$^{**}$ & $-$ 1.490 & $-$ 5.792$^{*}$ \\
  & ( 3.997) & ( 2.311) & ( 3.370) \\
Robust & $-$ 8.751$^{*}$ & $-$ 1.781 & $-$ 5.891$^{\dagger}$ \\
  & ( 4.726) & ( 2.902) & ( 3.995) \\
\addlinespace[2pt]
BW (km) &  62.8 &  68.6 &  80.7 \\
Eff.\ observations & 61 & 64 & 70 \\
\bottomrule
\end{tabular}
\begin{minipage}{0.92\textwidth}
\vspace{6pt}
\scriptsize
\textit{Notes:} This table reports two robustness checks for the nonparametric sharp RDD estimates of the effect of British colonial rule on Christian adherence. Panel~A adds eight geographic covariates (elevation, temperature, precipitation, caloric suitability, distance to coast, absolute latitude, terrain ruggedness, and distance to navigable river) to address the temperature imbalance documented in Table~1; the sample is the full 8 borders (122 observations). Panel~B restricts to the six British--French borders, excluding the two British--Portuguese segments (Zambia--Mozambique and Zimbabwe--Mozambique). All estimates use local linear regression with a triangular kernel and MSE-optimal bandwidth (Calonico, Cattaneo \& Titiunik 2014). \textit{Conventional} rows show the local linear point estimate with conventional standard errors; \textit{Robust} rows show the bias-corrected estimate with robust standard errors. 
$^{***}p<0.01$, $^{**}p<0.05$, $^{*}p<0.1$, $^{\dagger}p<0.15$.
\end{minipage}
\end{table}
}

\end{document}